\documentclass[aps,prb,twocolumn,groupedaddress, showpacs]{revtex4}
\usepackage[english]{babel}
\usepackage{amssymb,amsmath}
\usepackage{graphicx}
\usepackage{multirow}
\usepackage{bigstrut}

\begin{document}

\title{The cleavage surface of the BaFe$_{2-x}$Co$_{x}$As$_{2}$ and Fe$_{y}$Se$_{1-x}$Te$_{x}$ superconductors: \\from diversity to simplicity}

\author{F. Massee}\email{F.Massee@uva.nl}
\author{S. de Jong}
\author{Y. Huang}
\author{J. Kaas}
\author{E. van Heumen}
\author{J.B. Goedkoop}
\author{M.S. Golden}
\affiliation{Van der Waals-Zeeman Institute, University of Amsterdam, 1018XE Amsterdam, The Netherlands}

\date{\today}

\begin{abstract}
We elucidate the termination surface of cleaved single crystals of the BaFe$_{2-x}$Co$_{x}$As$_{2}$ and Fe$_{y}$Se$_{1-x}$Te$_{x}$ families of the high temperature iron based superconductors. By combining scanning tunneling microscopic data with low energy electron diffraction we prove that the termination layer of the Ba122 systems is a remnant of the Ba layer, which exhibits a complex diversity of ordered and disordered structures. The observed surface topographies and their accompanying superstructure reflections in electron diffraction depend on the cleavage temperature. In stark contrast, Fe$_{y}$Se$_{1-x}$Te$_{x}$ possesses only a single termination structure - that of the tetragonally ordered Se$_{1-x}$Te$_{x}$ layer.
\end{abstract}

\maketitle
\widowpenalty=300
\clubpenalty=300
Scanning tunneling microscopy and spectroscopy (STM/STS) and angle resolved photoemission (ARPES) are powerful and direct probes of the electronic states of solids and are making important contributions to our understanding of the new iron pnictide superconductors.\cite{Kamihara, STMreview, ARPESreview} Determination of the termination surface of the cleaved single crystals studied is a pre-requisite for meaningful application of STM/STS and ARPES, as is clarification whether possible departures from the bulk structure and co-ordination at the surface have a significant effect on the near-surface electronic states. A recent hard x-ray photoemission study \cite{Sanne} of room temperature cleaved BaFe$_{2}$As$_{2}$ (or Ba122) showed that the surface supports electronic states similar to those in the bulk, underpinning the relevance and reliability of surface sensitive techniques in the investigation of the pnictide superconductors. Low temperature STS of room temperature cleaved BaFe$_{1.86}$Co$_{0.14}$As$_{2}$ also showed no correlation between the spatial superconducting gap variation and the surface topography.\cite{massee}
However, studies on low temperature (\textless80K) cleaved samples report a correlation between features in the electronic spectra and the structure of the termination of the crystal,\cite{Boyer, yazdani} highlighting the role of the cleavage temperature. In the cuprate superconductors, we know that macroscopic departures from the simple lattice structure such as the incommensurate modulation in Bi$_{2}$Sr$_{2}$CaCu$_{2}$O$_{8+\delta}$, can have a pronounced effect on the spectroscopic data, leading to diffraction replicas in ARPES which can complicate matters considerably.\cite{Borisenko_Joys_Pitfalls} In the pnictide superconductors, the existence of diffraction replicas in ARPES data has already been pointed out.\cite{yazdani}

The first STM/STS studies of cleaved single crystals of the 122-based family of superconductors \cite{Boyer, Hoffman, yazdani, massee} either suggested or assumed that on average half of the alkaline earth ions (depending on the system involved) remain on each side of the cleave. Recent STM and low energy electron diffraction (LEED) studies, however, have concluded that the atomic contrast seen in STM of both Ba122 and Sr122 is from inequivalent As sites due to the spin density wave in the underlying Fe plane and that there is no cleavage temperature dependence.\cite{Plummer, Plummer2} Obviously, the fundamental importance of the surface termination question and the contrasting reports make it imperative to resolve these issues. Here, we present temperature dependent LEED analyses of cleaved BaFe$_{1.86}$Co$_{0.14}$As$_2$ single crystals, together with an extensive STM topographic database on this system obtained from cleavage both at low and high temperature. Comparison to the Fe$_{1-y}$Se$_{1-x}$Te$_{x}$ system which has the same Fe\textit{Pn} building block (where \textit{Pn} = As, Se or Te) as Ba122 but lacks the interstitial Ba layer, consequently enables us to isolate the Ba contribution to the cleavage surface in the 122 systems.

Single crystals of BaFe$_{1.86}$Co$_{0.14}$As$_{2}$ and Fe$_{1.07}$Se$_{0.45}$Te$_{0.55}$ were grown in a self flux, having superconducting T$_c$'s of 22$\pm$0.5K and 11$\pm$1K, respectively, determined from resistivity and AC susceptibility measurements. Cleavage took place both at low ($<$80K) and room temperature at a pressure better than 5$\times$10$^{-10}$ mbar directly before insertion into the STM head, where the samples were cooled to 4.2K. LEED was performed \textit{in situ} after each STM survey to obtain the azimuthal orientation of the crystal. The temperature dependent LEED experiments took place in a different vacuum system at a pressure of 7$\times$10$^{-11}$ mbar using identical crystals as those studied using STM, but cleaved at $<$ 25K. 

\begin{figure}[h]
	\centering
		\includegraphics[width=8cm]{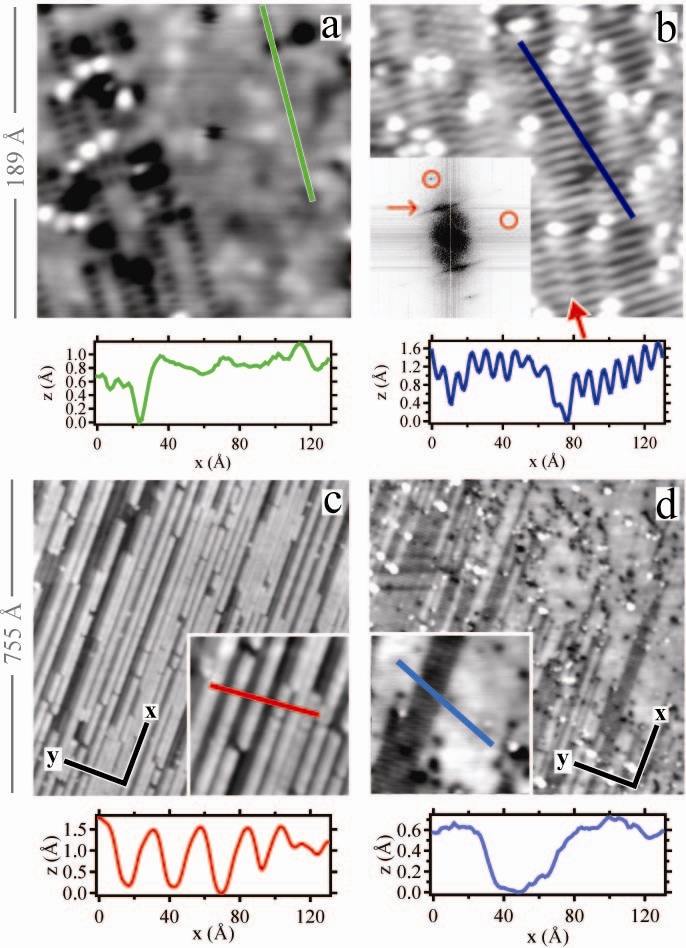}
	\caption{\label{fig:lowTtopo} Constant current images for cleavage of BaFe$_{1.86}$Co$_{0.14}$As$_{2}$ at low temperature (\textless80K). (a) and (b) are taken on the same cleave using a junction resistance R$_{J}$=2.4 G$\Omega$, (c) and (d) on a different cleave (R$_{J}$=0.65 G$\Omega$). All topographic types shown have been observed on numerous cleaves. Line scans along the paths indicated are shown below each panel. The insets of (c) and (d) show enlargements of size 189\AA.}
\end{figure}

Figure \ref{fig:lowTtopo} shows typical STM topographs obtained for low-T cleaves of BaFe$_{1.86}$Co$_{0.14}$As$_{2}$. In Fig. \ref{fig:lowTtopo}a the topography combines regions with very little contrast with stripe-like structures, the latter being very clear in Fig. \ref{fig:lowTtopo}b. This type of 2x1 or stripe feature has been observed previously both in the superconducting and parent 122 compound.\cite{Hoffman, Boyer, Plummer2} Furthermore, the data clearly show that the stripes possess phase shifts of half a unit (perpendicular to the stripes), which results in a ribcage-like topography, in which the phase shift lines correspond to the backbone (marked with an arrow in Fig. \ref{fig:lowTtopo}b). The inset shows a Fourier transform of the topographic image exhibiting dominant spots corresponding to the stripe/ribcage structures with a period of 8\AA\ (arrow) as well as weaker tetragonal unit cell spots (marked by circles). We note that the predominant stripe orientation is perfectly along the lattice and roughly perpendicular to the direction in which cleavage was performed. A third type of topographic situation is that shown in Fig. \ref{fig:lowTtopo}c: with larger ($\sim$20\AA\ wide), one dimensional rod-like features displaying a relatively large corrugation of approximately 1.5 \AA.
Completing our topographic survey of low-T cleaved crystals, Fig. \ref{fig:lowTtopo}d shows a large field of view in which two previously mentioned structures (seen in Fig. \ref{fig:lowTtopo}a and b) can be seen to smoothly cross over. Interestingly, Fig. \ref{fig:lowTtopo}a and d show that the backbone-features often terminate on pinning centers taking the form of a very bright or dark spot. Finally, we note that the linescan shown in Fig. 1d lacks steps corresponding to inter-plane distances (i.e. of order 1.4 \AA\ and larger), thus indicating that the majority of the observed structures form part of the same crystallographic plane.

\begin{figure}[h]
	\centering
		\includegraphics[width=8cm]{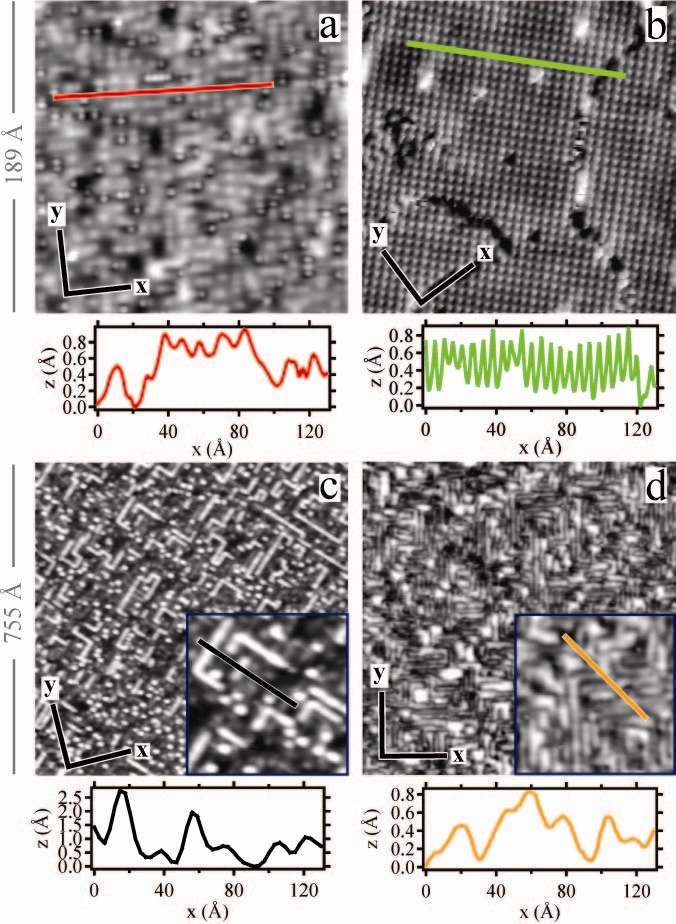}
	\caption{\label{fig:highTtopo} Constant current images for various cleaves of BaFe$_{1.86}$Co$_{0.14}$As$_{2}$ at room temperature (R$_{J}\sim$0.75 G$\Omega$). Line scans along the paths indicated are shown below each panel. The insets to (c) and (d) show enlargements of size 189\AA.}
\end{figure}

We now go on to show that room temperature cleavage results in quite a different set of constant current images of which four frequently occurring types are shown in Fig. \ref{fig:highTtopo}. The top left panel shows a highly disordered cleavage surface layer,\cite{massee} whereby the most frequent separation is $\sim$8\AA. Fig. \ref{fig:highTtopo}b shows a very clear $\sqrt{2}$x$\sqrt{2}$ topography with a period of 5.5\AA, cut by meandering anti-phase domain walls (`black rivers' in the topograph). Figure \ref{fig:highTtopo}c displays one of the more exotic cleavage surface structures encountered: a two dimensional network of stripes and dots, with a relatively large corrugation. Lastly, a diffuse maze-like landscape without atomic resolution, seen both on superconducting Ba122 and the parent compound,\cite{massee} is shown in Fig. \ref{fig:highTtopo}d and has also been reported in other studies.\cite{yazdani}

\begin{figure}[h]
	\centering
		\includegraphics[width=8cm]{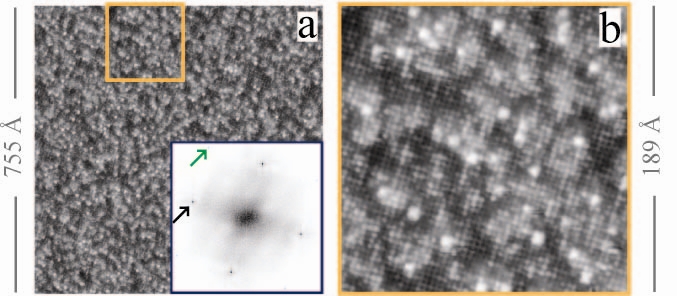}
	\caption{\label{fig:fesete}Constant current images of Fe$_{1.07}$Se$_{0.45}$Te$_{0.55}$, V$_{\text{sample}}$ = 20 mV, I$_{\text{set}}$ = 2 nA. (a) Large field of view with the inset showing the Fourier transform with spots corresponding to the atomic lattice vector (black arrow) and higher order spots (green arrow). (b) Zoom of the box in panel (a) showing the atomic resolution, visible even on the bright blobs.}
\end{figure}

We now turn our attention to Fe$_{y}$Se$_{1-x}$Te$_{x}$. For this system, both low and high temperature cleavage yield - without exception - identical results, in sharp contrast with the great variety of topographies observed in the Ba122 case. Fig. \ref{fig:fesete} shows a typical constant current image, in this case for a room-T cleave. Very clear atomic resolution is obtained over the entire field of view, with a total corrugation of less than 2\AA\ and a periodicity of 3.9\AA\, corresponding to the (Se,Te)-(Se,Te) distance. Clearly, no reconstruction is present on these surfaces, the simple tetragonal lattice being the only coherent structural pattern. The Fourier transform in the inset of Fig. \ref{fig:fesete}a emphasizes this: only tetragonal spots, also as higher harmonics, are observed. An interesting feature of the Fe$_{y}$Se$_{1-x}$Te$_{x}$ topographs is the occurrence of bright blobs on the field of view on top of which the atomic lattice is imaged. Their spatial abundance matches perfectly the excess Fe content (~7\%) in the crystals, and thus this seems a highly plausible explanation for these features. 

\begin{figure}[h]
	\centering
		\includegraphics[width=8cm]{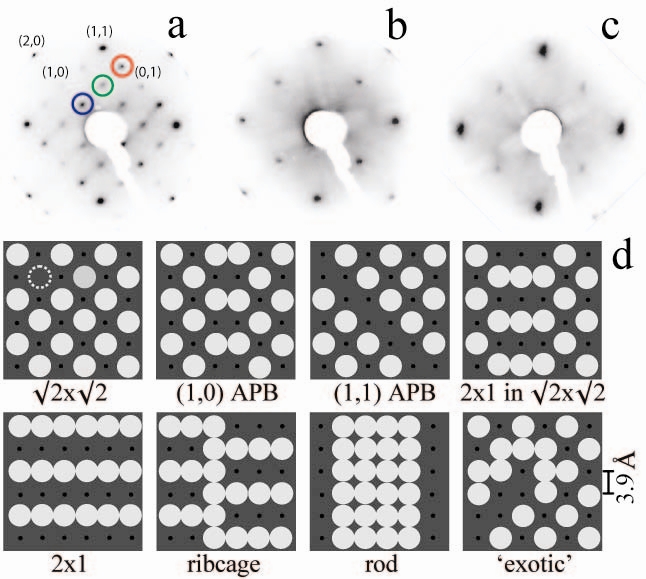}
	\caption{\label{fig:leed}Low energy electron diffraction, taken at 110eV. (a) BaFe$_{1.86}$Co$_{0.14}$As$_{2}$ cleaved and measured at 17K. The spots marked with a circle are the tetragonal ($\frac{1}{2}$,0) (blue), ($\frac{1}{2}$,$\frac{1}{2}$) (green) and ($\frac{1}{2}$,1) (red) reflections. (b) Same surface, but warmed up and measured at 200K. (c) Fe$_{y}$Se$_{1-x}$Te$_{x}$, cleaved and measured at 17K. (d) Sketch of various possible Ba surface overlayer configurations. Large (small) circles indicate the presence (absence) of a Ba atom. The dotted atom (left top) indicates a missing atom resulting in a black spot in topography. Extra atoms can similarly lead to bright spots. Anti-phase boundaries (APB) are shown running along two observed directions. Broader `backbones' in the ribcage structure are created by adding atoms at the phase shift boundary.} 
\end{figure}

To summarise the STM results, a large variety of cleavage surface termination structures is observed in the BaFe$_{1.86}$Co$_{0.14}$As$_{2}$ compound, which possess a strong cleavage temperature dependence. In contrast, the Fe$_{y}$Se$_{1-x}$Te$_{x}$ system, which lacks the interstitial Ba layer, has only a single type of topography, independent of cleavage temperature. 

Figure \ref{fig:leed} shows LEED data with E$_0$=110 eV. In Fig. \ref{fig:leed}a, the low temperature cleavage surface (17K) of the Ba122 superconductor gives rise to spots very clearly at positions other than those originating from the tetragonal unit cell of either Ba or As. 
Spots corresponding to both $\sqrt{2}$x$\sqrt{2}$ and 2x1 reconstructions in both a and b directions of the crystal are present. In the STM topographs of Fig. 1, we clearly image the 2x1 structures as the basic unit forming the ribcages. However, under our imaging conditions, the $\sqrt{2}$x$\sqrt{2}$ structures seen clearly in the high-T cleave do not generate sufficient contrast in the low-T cleaved data of Fig. 1a and b. For very low junction resistances, the $\sqrt{2}$x$\sqrt{2}$ topography has been imaged directly in the STM of low-T cleaved crystals of the parent compound Ba122 in Ref. [\onlinecite{Plummer}]. 
Returning to the LEED data, upon increasing the temperature, all the non-tetragonal LEED spots start to lose intensity above 100K, and by reaching 200K, only the tetragonal reflections remain (Fig.\ref{fig:leed}b). More importantly, on re-cooling the sample back to 17K, \textit{none} of the non-tetragonal spots reappear. The main spots, however, become more diffuse once the cleave is or has been at higher temperature, signalling increased surface disorder or the presence of structurally incoherent reconstructions.
Having presented the LEED data for BaFe$_{1.86}$Co$_{0.14}$As$_{2}$, we now show a typical LEED pattern from the Fe$_{y}$Se$_{1-x}$Te$_{x}$ compound cleaved and measured at low temperature in Fig. \ref{fig:leed}c. Not surprisingly, since the STM images only show the tetragonal atomic lattice, none of the extra LEED spots appearing in the Ba122 case are present in the Fe$_{y}$Se$_{1-x}$Te$_{x}$ data. 

Returning to the LEED data from the iron arsenide system, tracking the intensity of the main (tetragonal) spots seen in Fig. \ref{fig:leed}a and b results in I(V) curves which follow the theoretical curves for diffraction from the As layer reported in Fig. 4a of Ref. [\onlinecite{Plummer}]. This agreement between our data with LEED theory confirms that the outermost \textit{complete} layer after cleavage consists of As atoms.\cite{Plummer} However, and importantly, the LEED data from the low temperature cleaves of BaFe$_{1.86}$Co$_{0.14}$As$_{2}$ and the parent compound BaFe$_{2}$As$_{2}$ (data not shown) clearly show ($\frac{1}{2}$,$\frac{1}{2}$)$_{\text{tetragonal}}$ spots, in contrast to the  data of Ref. [\onlinecite{Plummer}]. The I(V) trace of our ($\frac{1}{2}$,$\frac{1}{2}$)$_{\text{tetragonal}}$ spots is very different to both the experimental and theoretical I(V) curves in Ref. [\onlinecite{Plummer}], possessing peaks and minima at totally different energy values. The mismatch between our experiment and the theory in Ref. [\onlinecite{Plummer}] rules out the As layer as the origin of our extra LEED spots, leaving the partially ordered Ba atom overlayer as the obvious candidate to explain the ($\frac{1}{2}$,$\frac{1}{2}$)$_{\text{tetragonal}}$ spots in LEED and the topographic features seen in STM with periods greater than the tetragonal repeat unit of 3.9 \AA. 
A surprising observation in the STM topograph of Fig. \ref{fig:highTtopo}b is the clear $\sqrt{2}$x$\sqrt{2}$ lattice on the surface of a high temperature cleaved sample, whereas the corresponding ($\frac{1}{2}$,$\frac{1}{2}$)$_{\text{tetragonal}}$ spots in LEED have ceased to be observable above the background upon warming up the sample. At this point, we mention that theoretical calculations predict that the $\sqrt{2}$x$\sqrt{2}$ (or checkerboard) reconstructed surface is the most energetically favourable ordered distribution of the remaining half of the Ba atoms on either side of the cleave.\cite{kelly}
One approach to understand why the $\sqrt{2}$x$\sqrt{2}$ LEED spots disappear at higher temperature is offered by consideration of the dislocation networks seen in the STM topographic data in the high temperature cleaved surfaces (the `black rivers' seen in Fig. 2b), which are absent in the low-T cleave topographs. Upon warm cleavage or increase of the sample temperature above ca. 200K, it is conceivable that fragmentation of the checkerboard order into nanoscopic domains causes the coherence length of the reconstruction to drop sufficiently such that the $\sqrt{2}$x$\sqrt{2}$ spots fade into the background of the LEED pattern, which itself is the result of a measurement which spatially integrates on a mm length-scale.

By now it is clear that the topographical diversity of the cleavage surface of the alkaline earth-122 material family is due to the alkaline earth atom overlayer that remains after cleavage. In fact, we can reconcile all our experimental data from the 122 systems within a very simple structural model, based only on differences in the distribution of the Ba atoms within the overlayer that remains on the surface after cleavage.
A sketch of the model is shown in Fig. \ref{fig:leed}d. When a sample is cleaved at low temperature, the Ba surface atoms are not mobile enough to re-arrange themselves into the energetically most favourable structure with metastable situations such as the 2x1, or rib-cage reconstruction as the result. Increasing temperature enables the overlayer atoms to re-arrange, thus patches of checkerboard reconstruction are formed separated by domain walls (`black rivers'). Alternatively, more disordered configurations such as those seen in Fig. 2a are also possible, depending on the local concentration of Ba atoms.
Ordering of the Ba surface atoms in locally densely packed structures surrounded by a relatively Ba-free region could lead to an increased corrugation in STM, matching the rods seen in Fig. \ref{fig:lowTtopo} and the bright regions in Fig. \ref{fig:highTtopo}c. However, these kind of structures are still within the same Ba layer as the locally less densely packed ordered $\sqrt{2}$x$\sqrt{2}$ and 2x1 regions, and thus these different topographies can cross over into one another as indeed has been observed. Evidently, more random distribution of the Ba atoms in the overlayer could lead to topographs as in Fig. \ref{fig:highTtopo}a and d. 

In summary, Ba122 exhibits a large variety of topographies and has pronounced cleavage temperature dependence. In LEED, spots corresponding to large scale coherent structures of dimensions larger than the 3.9 \AA\ tetragonal unit cell, i.e. 2x1 and $\sqrt{2}$x$\sqrt{2}$, appear for low temperature cleaved samples. These spots vanish upon warming the sample to 200K and do not reappear after cooling back to 17K. Since superconducting BaFe$_{1.86}$Co$_{0.14}$As$_{2}$ possesses no spin density wave at 4.5K,\cite{Laplace} the 2x1 and $\sqrt{2}$x$\sqrt{2}$ topographies in STM cannot be explained by a SDW-driven creation of two distinct As sites, but only by the ordering of the Ba atoms remaining on the surface after cleavage. We propose that the disappearance of the extra spots seen in LEED at higher temperatures is a signal of fragmentation of these structures due to a re-distribution of the Ba atoms at the surface into energetically more favourable configurations on a small length scale. These still appear as ordered patterns in STM, yet possess insufficient long range order to support LEED spots. The longer range order does not re-establish itself upon re-cooling, thus the non-tetragonal spots do not reappear and the tetragonal LEED spots themselves become more diffuse. Our argument that the non-tetragonal LEED spots originate from the Ba termination layer is supported by the fact that our experimental LEED I(V) profiles of the tetragonal ($\frac{1}{2}$,$\frac{1}{2}$) spots disagree totally with LEED I(V) calculations in which these features are taken as diffraction from the arsenic plane.
In stark contrast to the richness and complexity of the surface topography of the Ba122 family of compounds (and by extrapolation to the other \textit{M}122 systems, \textit{M}=Ba, Eu, Sr...), Fe$_{y}$Se$_{1-x}$Te$_{x}$ displays only one type of cleavage surface and LEED pattern without any sign of reconstructions, independent of the cleavage temperature.

We would like to acknowledge H. Agema and H. Luigjes for expert technical support and C.C. Tsui, W. Plummer and V.B. Nascimento for useful discussions. This work is part of the research programme of FOM, which is financially supported by the NWO.

\end{document}